\documentclass{jps-cp}
\usepackage{txfonts} %Please comment out this line unless the txfonts package is availabe in your LaTeX system.
\usepackage{graphicx}

\title{Quark mass effects in the thermodynamical properties of an extended (P)NJL model}
\author{J. \textsc{Moreira}$^{1}$, J. \textsc{Morais}$^{1}$, B. \textsc{Hiller}$^{1}$,  A. H. \textsc{Blin}$^{1}$ and A. A. \textsc{Osipov}$^{2}$}
\inst{
$^{1}$CFisUC, Department of Physics, University of Coimbra, P-3004-516 Coimbra, Portugal \\
$^{2}$Bogoliubov Laboratory of Theoretical Physics, JINR, Dubna, Moscow Region,Russia
}
\email{jmoreira@uc.pt}
\recdate{\today}
\abst{We analyze the thermodynamical properties of a system of strongly interacting particles at vanishing quark chemical potential in the framework of a recently developed extension of the Polyakov-Nambu-Jona-Lasinio Model. In addition to eight quark interactions terms, non-canonical terms which explicitly break chiral symmetry up to the same order in a $1/N_c$ expansion ($N_c$ number of colors) are included. A recently proposed Polyakov potential is considered and the results are compared to lattice QCD data resulting in a favorable scenario  for the recent model variants.}
\kword{spontaneous chiral symmetry breaking, explicit chiral symmetry breaking, general spin 0 eight-quark interactions, finite temperature}
\begin{document}
\maketitle
\section{Introduction}
Simple effective models are useful tools in the study of systems of strongly interacting particles. Among these, models of the Nambu-Jona-Lasinio type 
%\cite{Nambu:1961tp,Nambu:1961fr}
\cite{NJL}, drawing its power from the fact that they share with QCD its global symmetries and include both spontaneous and explicit chiral symmetry breaking, have been quite successful. The fact that they are not plagued by the sign problem of the \emph{ab initio} approach of lattice QCD (lQCD) at finite quark chemical potential ($\mu_q$) has a twofold consequence: on the one hand this means that their degree of success can be checked by comparison to the result of lQCD simulations at $\mu_q=0$ and on the other hand they can be used to venture into the region of the phase diagram  which is problematic to lQCD.
\subsection{The model}
Restricting our study to the light quark sector ($u$, $d$ and $s$), a recently developed model lagrangian (NJLH8qm) 
%\cite{Osipov:2012kk,Osipov:2013fka}
\cite{NJLH8qm} includes terms with explicit chiral symmetry breaking up to the same order in a $1/N_c$ expansion as the 't Hooft determinant term ($6q$ interaction term) \cite{Hooft,Kunihiro:1987bb,Bernard:1987sg,Reinhardt:1988xu} which one must add to the NJL interaction (the $4q$ interaction term which drives the spontaneous breaking of chiral symmetry) to break the unwanted $U_A(1)$ axial symmetry. This is an extension of a previously developed version with $8q$ terms (NJLH8q)\cite{Osipov:2005tq,Osipov:2006ns} and consistently contains all non-derivative spin 0 multiquark interactions up to the the relevant order in $1/N_c$.

Polyakov loop extension of these models 
\cite{Fukushima:2003fw,
% Megias:2003ui,Megias:2004hj,
Roessner:2006xn,
%Ghosh:2007wy,Fu:2007xc,Costa:2008dp,
Fukushima:2008wg,
%Bhattacharyya:2010wp,
Moreira:2010bx%,Stiele:2016cfs,Bhattacharyya:2016jsn
}
 has become a very popular way to mimic the inclusion of gluonic degrees of freedom. In these models an additional Polyakov potential must be introduced to drive the deconfinement transition (the Polyakov loop goes from $\phi=~0\rightarrow 1$). Several forms of this potential have been considered trying to reproduce lQCD results at vanishing $\mu_q$. Here we will consider the recently proposed from  \cite{Ghosh:2007wy,Bhattacharyya:2016jsn} and compare lQCD results  reported in 
\cite{
Borsanyi:2011sw,
Bazavov:2012jq,
Borsanyi:2013bia,
Bazavov:2014pvz
}
.

A more detailed description of the model can be found in \cite{Moreira:2018xsp} where the results without the Polyakov loop extension and the results considering the Polyakov potential form from \cite{Roessner:2006xn} are also considered. 

\section{Results and discussion}

Here we will analyze some thermodynamical quantities resulting from the Polyakov extended version of the model using four different parameter sets: two in the 
NJLH8q model and two in the NJLH8qm model. The sets are labeled with suffix $A/B$ referring to weaker/stronger OZI-violating $8q$ interactions.
% although it should be mentioned that in the case of NJLH8qm a realistic fit to the meson spectra locks us in a moderately strong 8q interaction regime when compared to the NJLH8q case. 
%
It should be noted that, while in the NJLH8q model a wide range of fits differing in strengths of the OZI-violating $8q$ interactions is possible (an increase in the $8q$ interaction strength accompanied by a suitable decrease in the NJL $4q$ interaction leaves the meson spectra unchanged apart from a decrease in the $\sigma$ meson mass) \cite{Osipov:2006ns}, in the case of NJLH8qm a much more realistic fit to the meson spectra is achieved while at the same time locking us in a moderately strong 8q interaction regime when compared to the NJLH8q case thus reducing the arbitrariness in the $4q$/$8q$ interactions interplay.

\begin{figure}
\centering
	\includegraphics[width= 0.33 \textwidth]{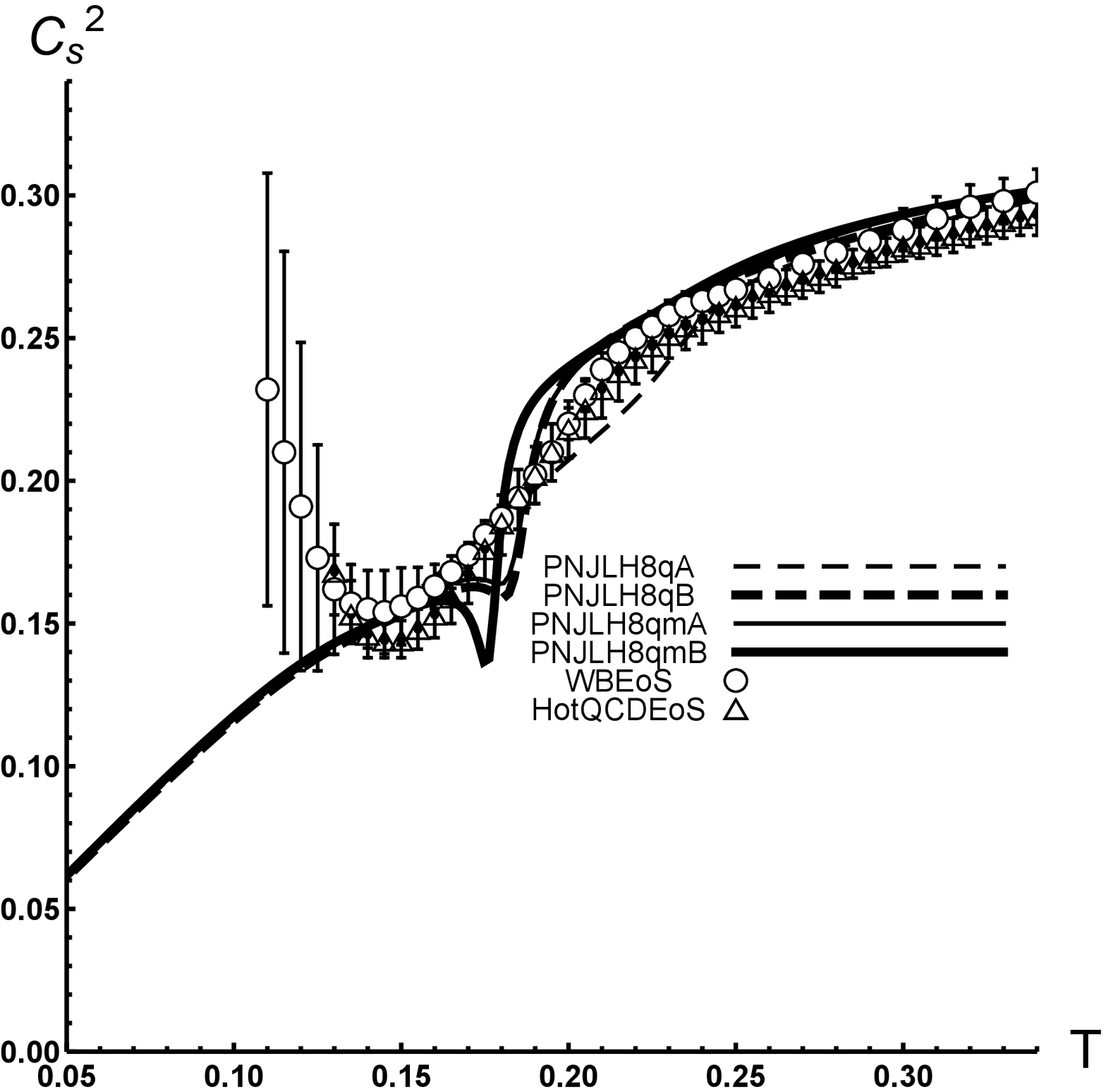}\hfill
	\includegraphics[width= 0.33 \textwidth]{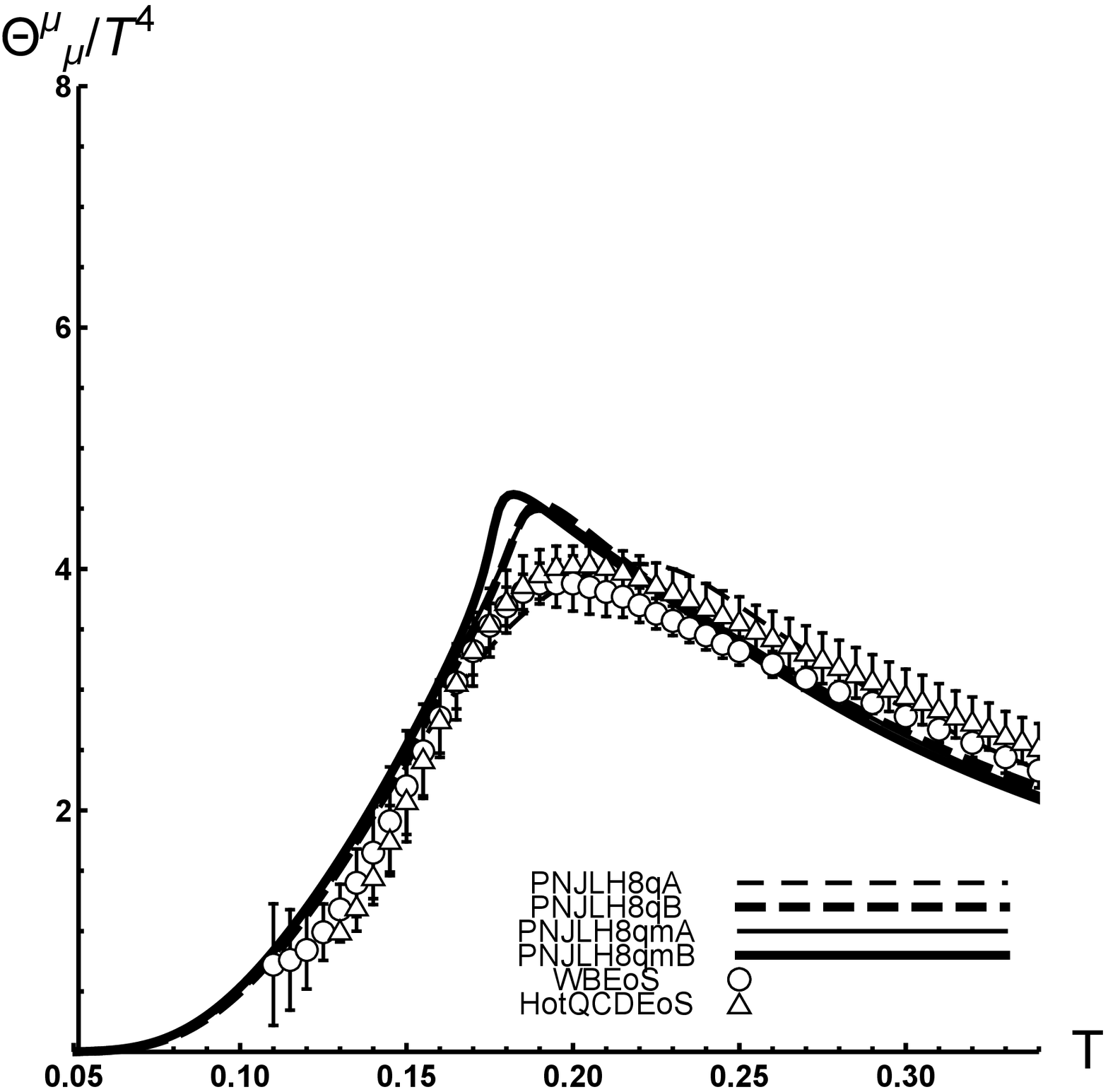}\hfill
	\includegraphics[width= 0.33 \textwidth]{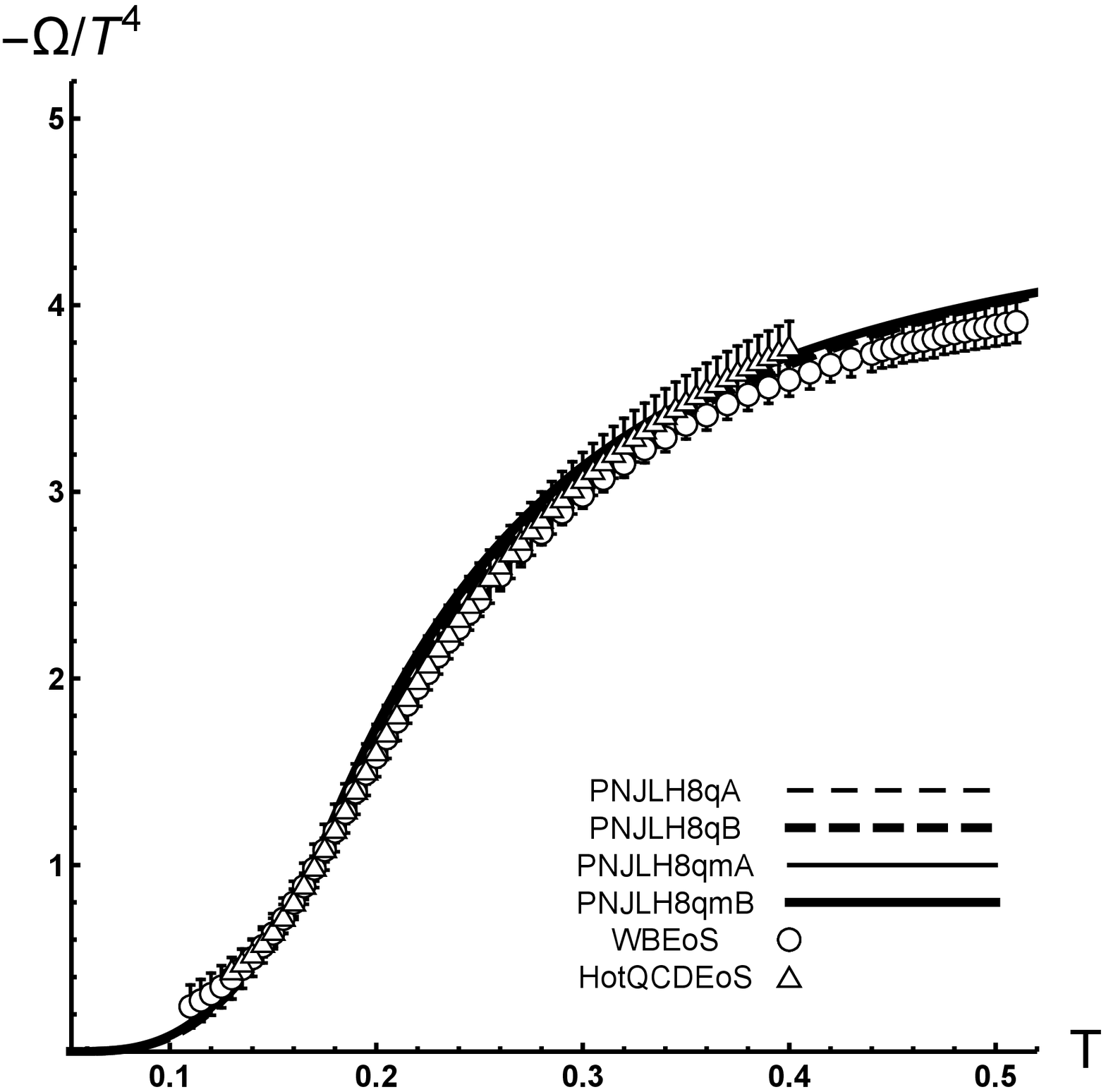}\\
	\includegraphics[width= 0.33 \textwidth]{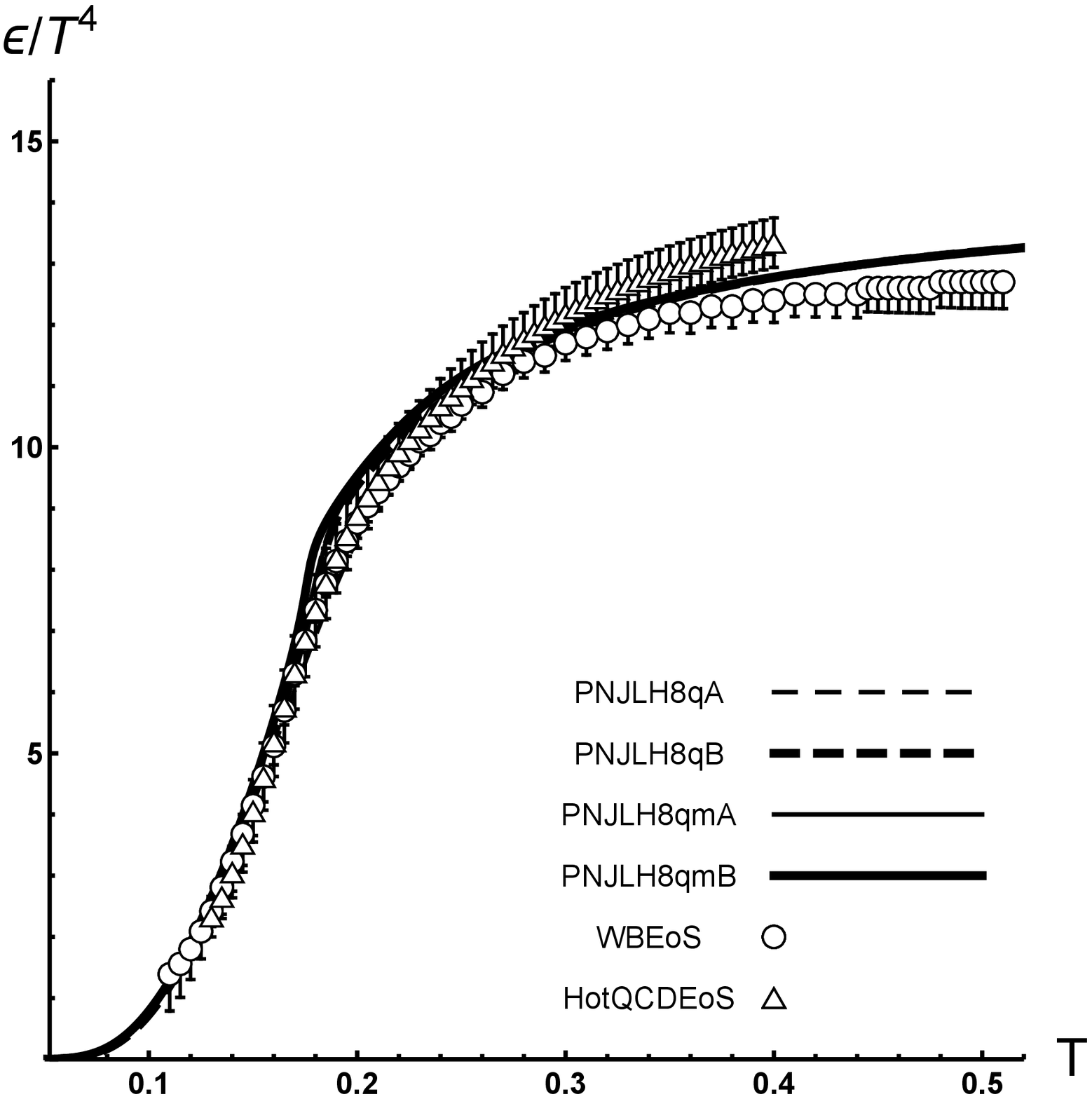}\hfill
	\includegraphics[width= 0.33 \textwidth]{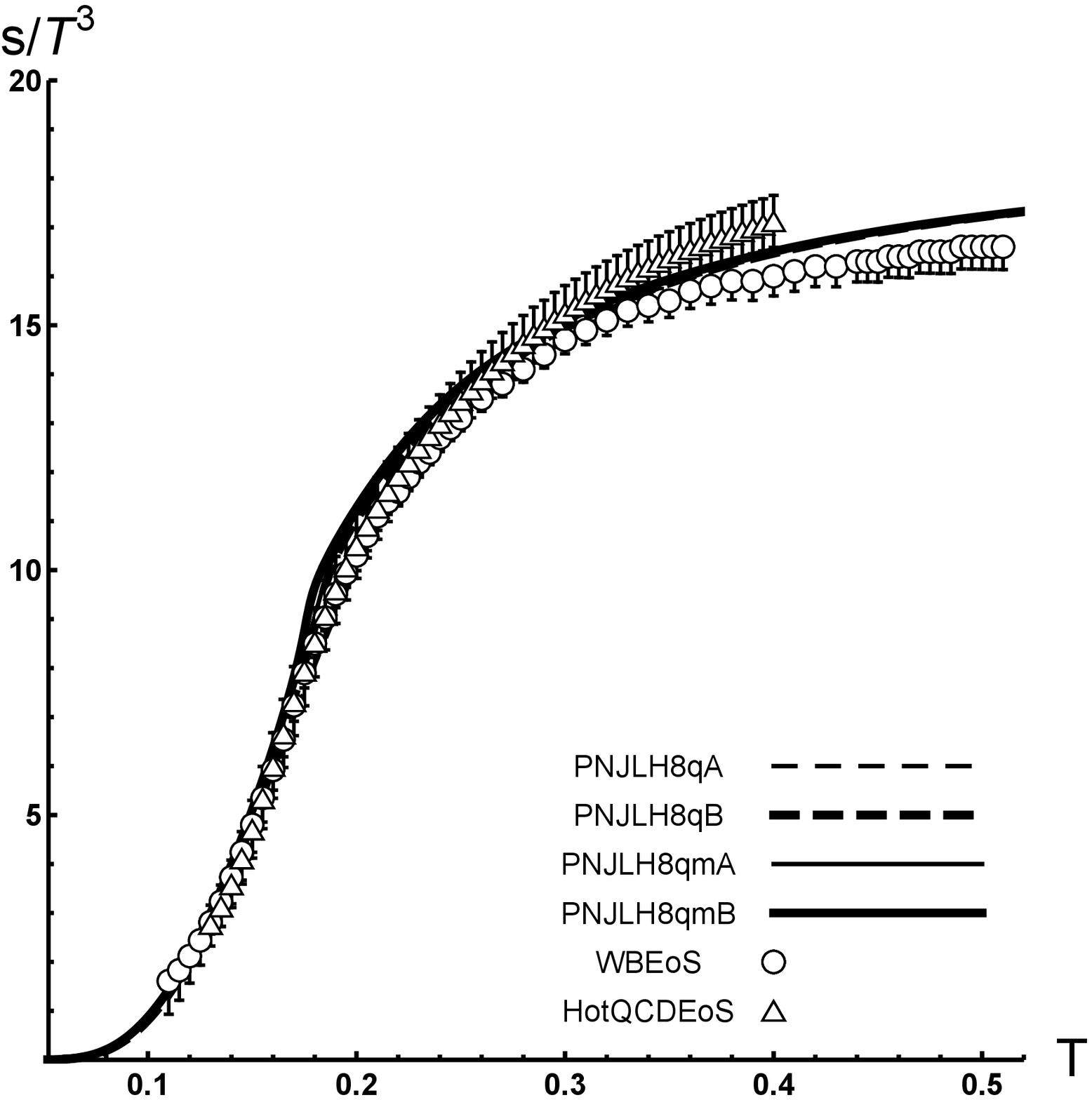}\hfill
	\includegraphics[width= 0.33 \textwidth]{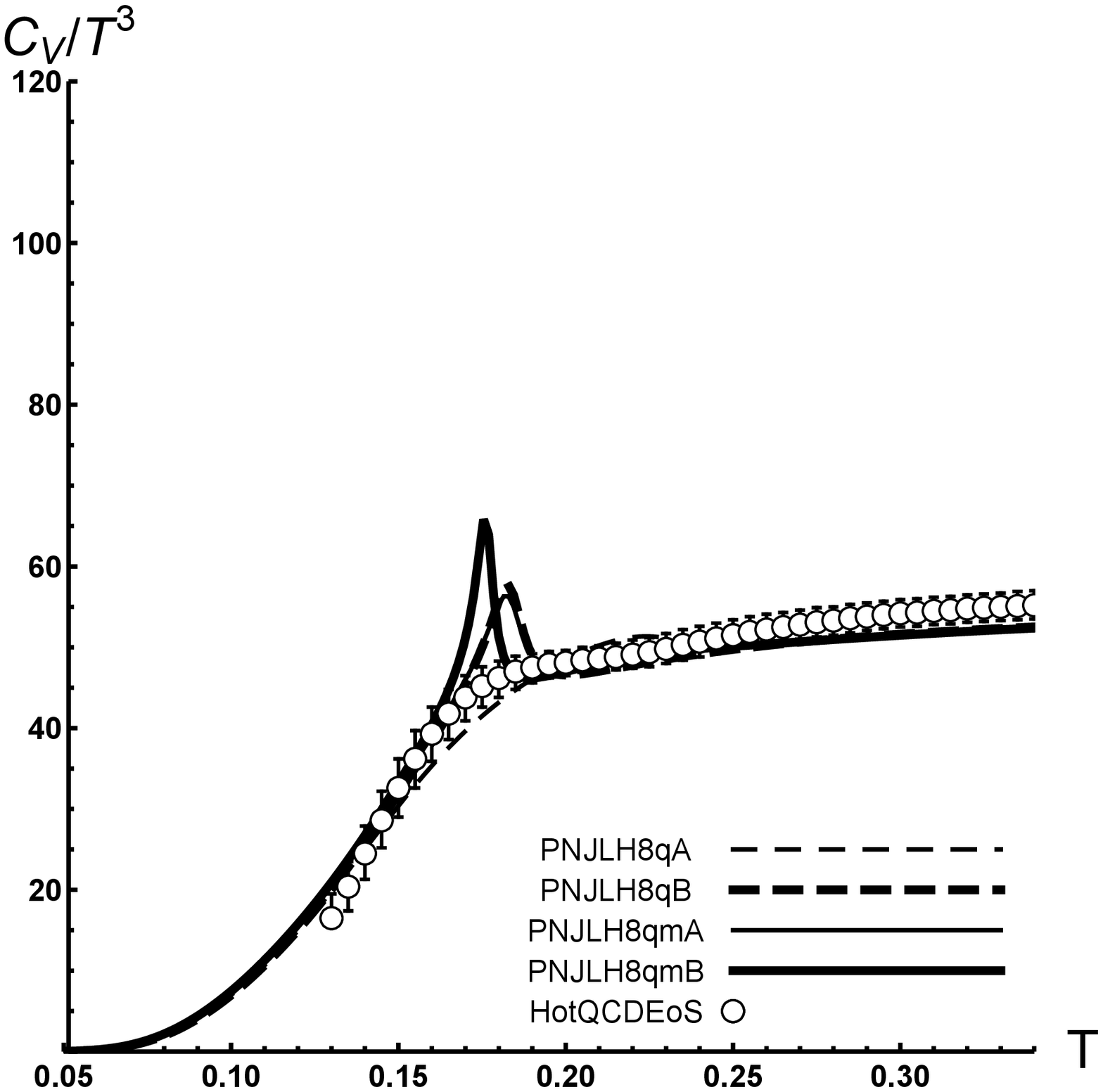}\hfill
\caption{
From left-to-right, top-to-bottom: $C_s^2$, $\Theta^\mu_\mu$, $-\Omega$, $\epsilon$, $s$ and $C_V$ as functions of the temperature ($[T]=\mathrm{GeV}$) at $\mu_q=0$, obtained using parameter sets with stronger and weaker OZI-violating $8q$ interactions ($A/B$ refers to the weak/strong case respectively): dashed lines correspond to the NJLH8q sets whereas solid lines correspond to the NJLH8qm sets. The markers, labeled as WBEoS and HotQCDEoS, correspond to continuum extrapolated lQCD results taken respectively from \cite{Borsanyi:2013bia} and \cite{Bazavov:2014pvz}.
}
\label{Thermo_PELAPV}
\end{figure}

\begin{figure}
\centering
    \includegraphics[width=0.33\textwidth]{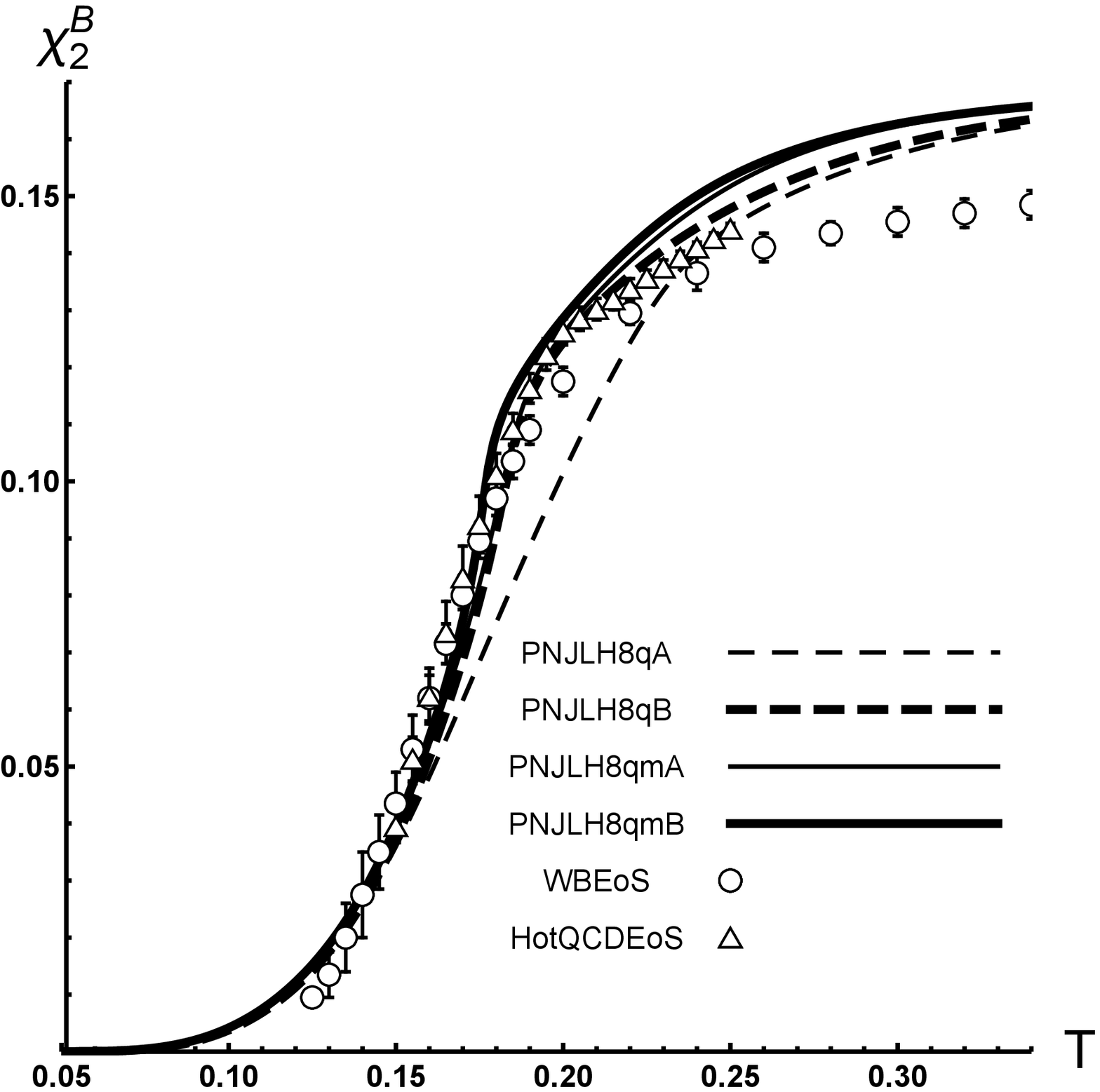}\hfill
    \includegraphics[width=0.33\textwidth]{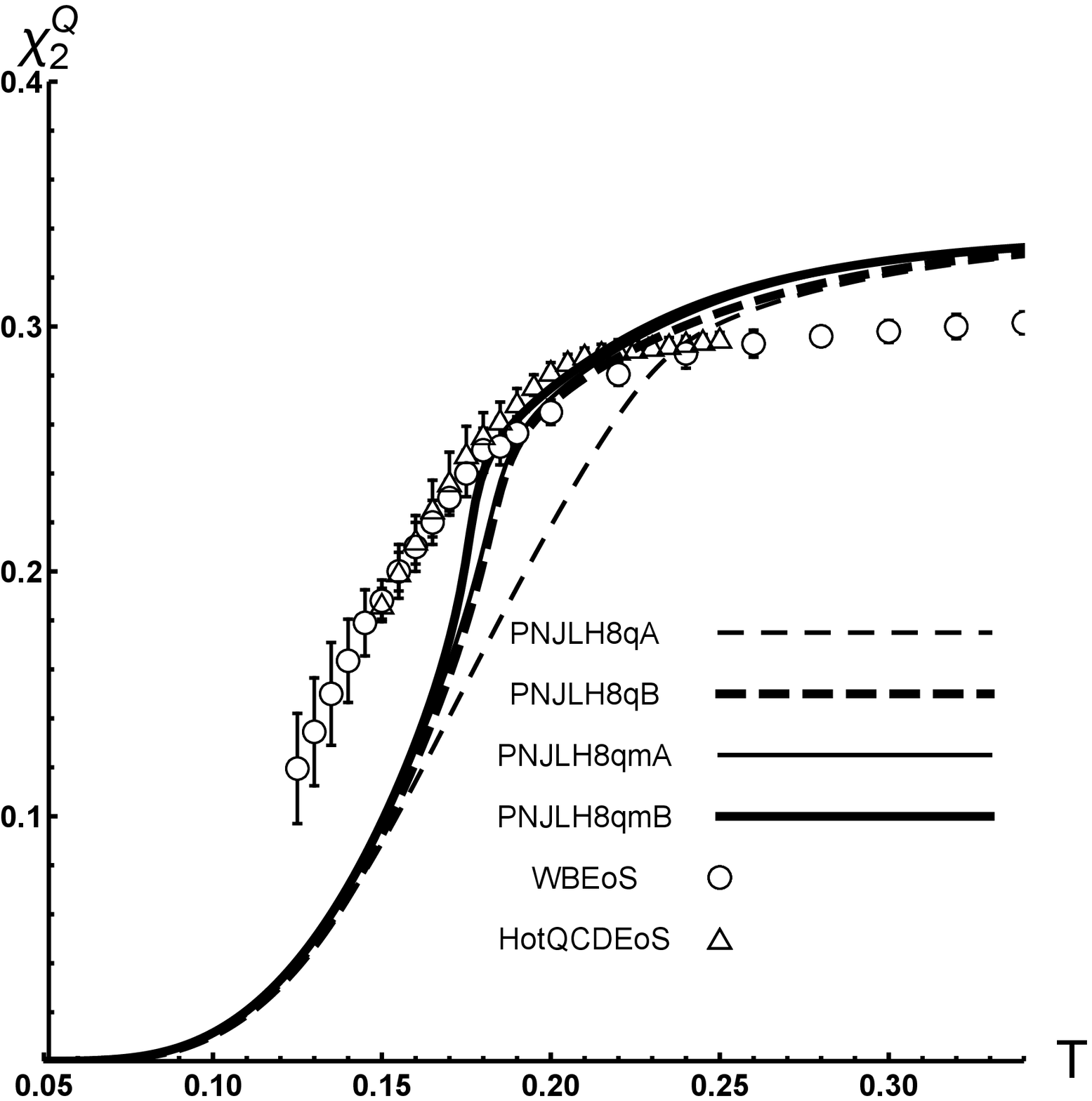}\hfill
		\includegraphics[width=0.33\textwidth]{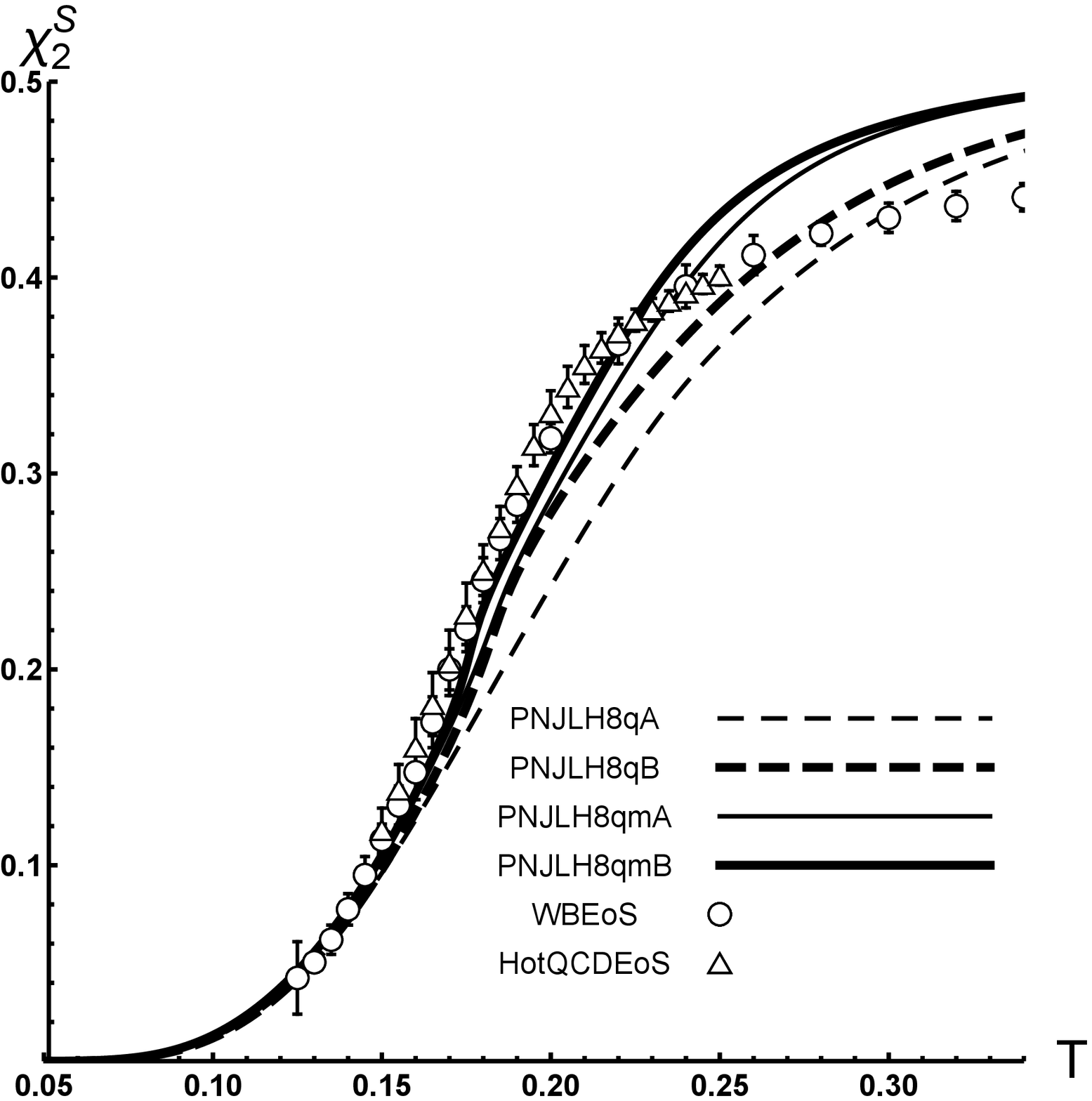}\\
%\caption{asdnabsnd}
%\label{Chi2BPELAPV}
%\end{figure}
%\begin{figure}
%\centering
	\includegraphics[width= 0.33 \textwidth]{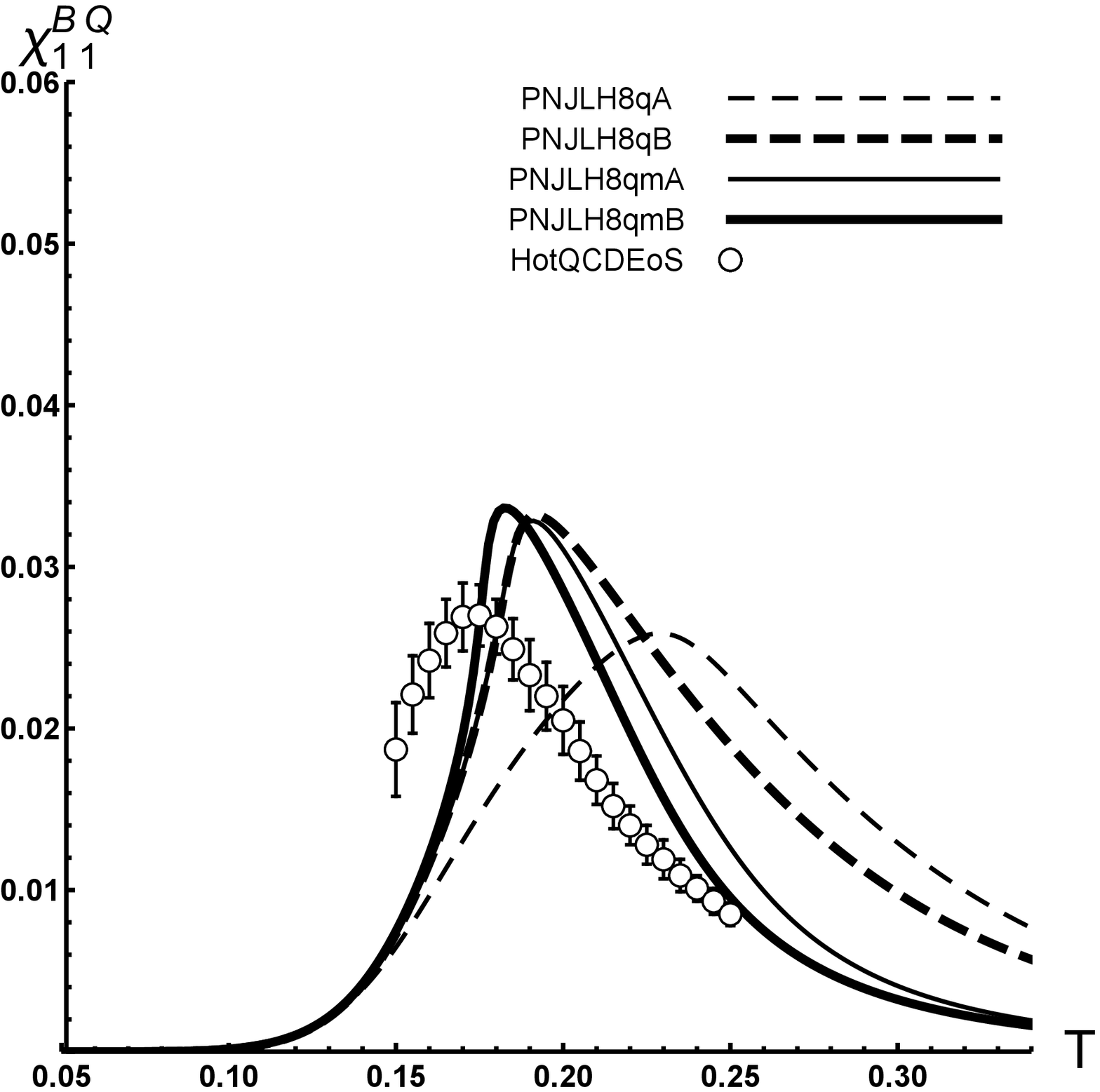}\hfill
	\includegraphics[width= 0.33 \textwidth]{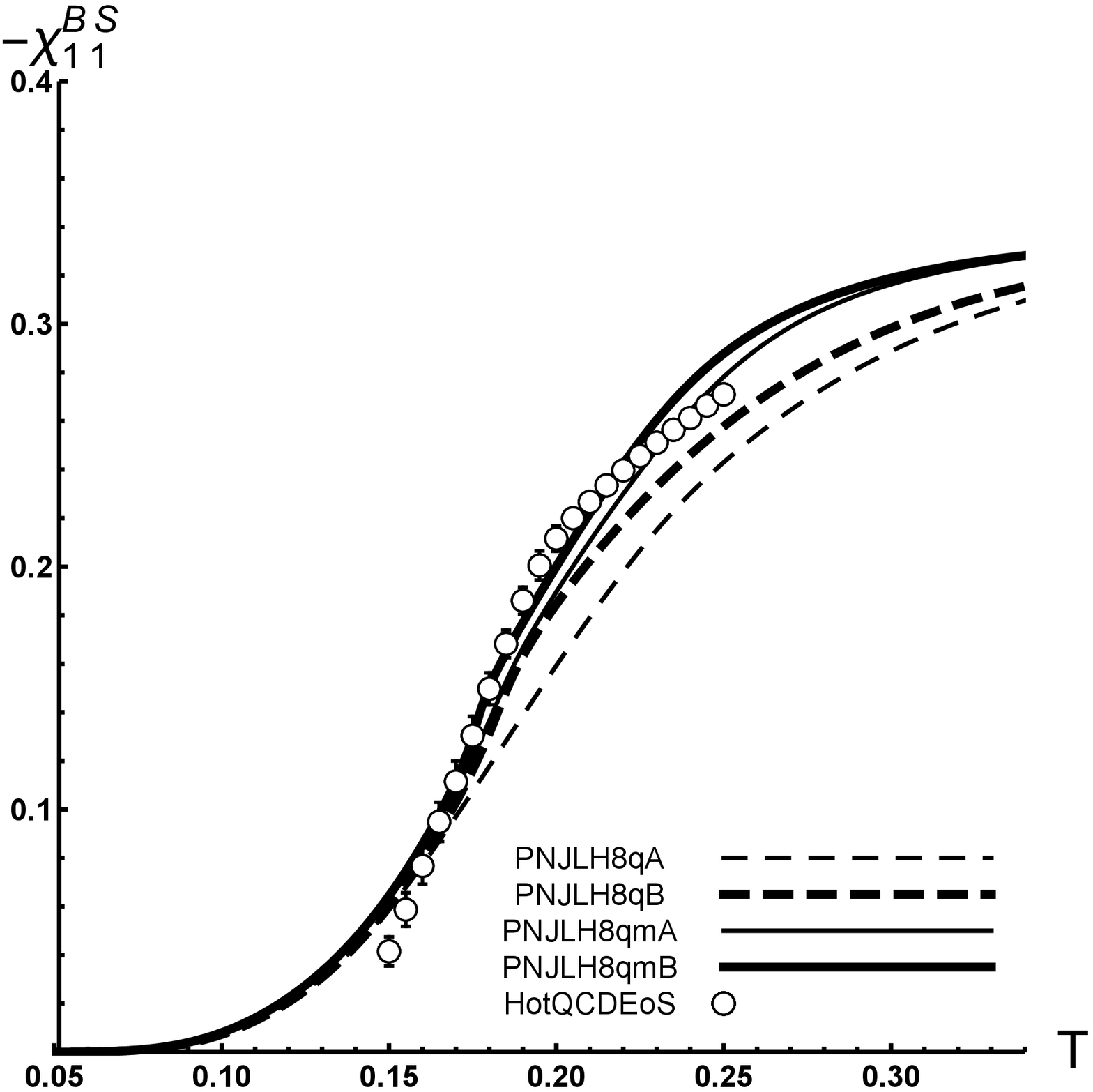}\hfill
	\includegraphics[width= 0.33 \textwidth]{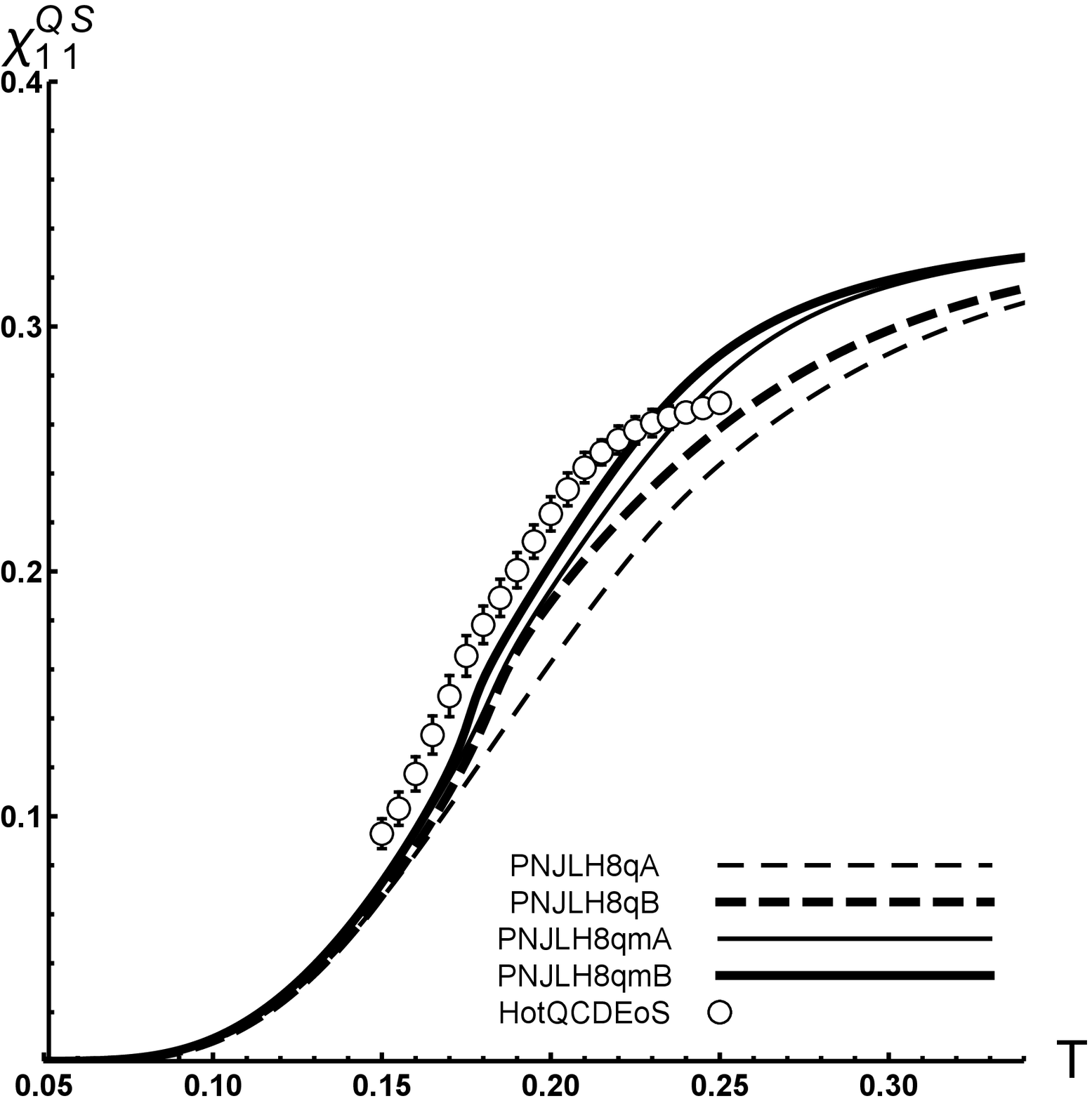}
\caption{Top row, left-to-right displays the fluctuations of conserved charges: $\chi^B_2$, $\chi^Q_2$ and $\chi^S_2$. Bottom row, left-to-right, correlations: 
$\chi^{BQ}_{11}$, $\chi^{BS}_{11}$ and $\chi^{QS}_{11}$. Same model parameters and notation as in Fig. \ref{Thermo_PELAPV}. The markers, labeled as WBEoS and HotQCDEoS, correspond to continuum extrapolated lQCD results taken respectively from \cite{Borsanyi:2011sw} and \cite{Bazavov:2012jq}.}
\label{Fluct_Correl_PELAPV}
\end{figure}

In Fig. \ref{Thermo_PELAPV} we present the $T$ dependence at $\mu_q=0$ of several thermodynamical quantities: squared speed of sound ($C_s^2$), the energy-momentum tensor trace anomaly ($\Theta^\mu_\mu$), pressure ($-\Omega$), energy density ($\epsilon$), entropy density ($s$) and specific heat 
($C_V$). All these quantities display a marked behavior showing the crossover transition which according to recent lQCD data is expected to occur at a temperature $T\approx 155~\mathrm{MeV}$ over a range of $\Delta_T\approx 20 \mathrm{MeV}$ \cite{Borsanyi:2013bia, Bazavov:2014pvz}.
%and compared to lQCD results. 

In Fig. \ref{Fluct_Correl_PELAPV} the fluctuations and correlations of conserved charges (baryon number, electric charge and strangeness) can be observed. These are thought to be good probes for the experimental detection of the transition behavior in relativistic heavy-ion collisions. Note that for that application one should consider finite $\mu_q$ but here we are looking for the validation of the model against lQCD data at $\mu_q=0$. 

As can be seen in Figs. \ref{Thermo_PELAPV} and \ref{Fluct_Correl_PELAPV} there is a good general agreement with lQCD simulations. Several points are worthy of note:
\begin{itemize}
\item the stronger eight quark interactions appear to be necessary for the reproduction of the local minimum of $C_s^2$;
\item $\Theta^\mu_\mu$, $-\Omega$, $\epsilon$ and $s$ are all reasonably reproduced;
\item $C_V$ displays a peak (except for the NJLH8qA case) which is absent from the lQCD data which reflects the overall faster transition behavior induced by the OZI-violating 8q interactions;
\item baryon number and strangeness fluctuations ($\chi^B_2$ and $\chi^S_2$) as well as the baryon number-electric charge, baryon number-strangeness and electric charge-strangeness correlations ($\chi^{BQ}_{11}$, $\chi^{BS}_{11}$ and $\chi^{QS}_{11}$) are reasonably reproduced and have their slope improved by the stronger 8q interaction regime and the inclusion of non-canonical explicit chiral symmetry breaking interactions;
\item the electric charge fluctuation $\chi^B_2$ is not accurately reproduced in our model.
\end{itemize}

\section{Conclusions}

Building upon the success of the recently proposed NJLH8qm model in reproducing the low-lying spectra of scalar and pseudoscalar mesons, it is shown that
its Polyakov loop extended version is quite successful in reproducing lQCD results for several relevant thermodynamical quantities. This strengthens the importance of this model as a simple tool for the study of the thermodynamics of systems of strongly interacting particles at finite $T$ and $\mu$ for instance in the study of the fluctuations and correlations of conserved charges in regions of the phase diagram relevant to relativistic heavy ion collisions.

\section*{Acnowledgments}
Research supported by CFisUC and Fundac\~ao para a Ci\^encia e Tecnologia 
through the project UID/FIS/04564/2016,  and grants SFRH/BD/110315/2015,  SFRH/BPD/110421/2015.
We acknowledge networking support by the COST Action CA16201.

%\bibliography{JMoreiraProceedingsQNP2018}

\begin{thebibliography}{9}
\bibitem{NJL} Y. Nambu and G. Jona-Lasinio, Phys. Rev. \textbf{122}, 345(1961); Phys. Rev. \textbf{124}, 246(1961).
%\bibitem{Nambu:1961tp}  Y. Nambu and G. Jona-Lasinio, Phys. Rev. \textbf{122}, 345(1961).
%\bibitem{Nambu:1961fr}  Y. Nambu and G. Jona-Lasinio, Phys. Rev. \textbf{124}, 246(1961).
\bibitem{NJLH8qm} A. Osipov, B. Hiller, and A. Blin, Eur.Phys.J. A\textbf{49}, 14 (2013);  Phys.Rev. D\textbf{88}, 054032 (2013).
%\bibitem{Osipov:2012kk} A. Osipov, B. Hiller, and A. Blin, Eur.Phys.J. A\textbf{49}, 14 (2013).
%\bibitem{Osipov:2013fka} A. Osipov, B. Hiller, and A. Blin, Phys.Rev. D\textbf{88}, 054032 (2013).
\bibitem{Hooft} G. 't Hooft, Phys. Rev. D\textbf{14}, 3432 (1976); Phys. Rev. D\textbf{18}, 2199 (1978).
%\bibitem{'tHooft:1976fv} G. 't Hooft, Phys. Rev. D\textbf{14}, 3432 (1976)
%\bibitem{PhysRevD.18.2199.3} G. 't Hooft, Phys. Rev. D\textbf{18}, 2199 (1978).
\bibitem{Kunihiro:1987bb} T. Kunihiro and T. Hatsuda, Phys.Lett. B\textbf{206}, 385 (1988).
\bibitem{Bernard:1987sg} V. Bernard, R. L. Jaffe, and U. G. Meissner, Nucl. Phys.B\textbf{308}, 753 (1988).
\bibitem{Reinhardt:1988xu} H. Reinhardt and R. Alkofer, Phys.Lett. B\textbf{207}, 482 (1988).
\bibitem{Osipov:2005tq} A. A. Osipov, B. Hiller, and J. da Providencia, Phys. Lett. B\textbf{634}, 48 (2006).
\bibitem{Osipov:2006ns}  A. A. Osipov, B. Hiller, A. H. Blin, and J. da Providencia, Annals Phys. \textbf{322}, 2021 (2007).
%!!!AGRUPAR?
%\bibitem{Nambu:1961tp}  Y. Nambu and G. Jona-Lasinio, Phys. Rev. \textbf{122}, 345(1961); Phys. Rev. \textbf{124}, 246(1961)
\bibitem{Fukushima:2003fw} K. Fukushima, Phys. Lett. B\textbf{591}, 277 (2004)
%\bibitem{Megias:2003ui} E. Megias, E. Ruiz Arriola, and L. L. Salcedo, Phys. Rev. D\textbf{69}, 116003 (2004).
%\bibitem{Megias:2004hj} E. Megias, E. Ruiz Arriola, and L. Salcedo, Phys.Rev. D\textbf{74}, 065005 (2006).
\bibitem{Roessner:2006xn} S. Roessner, C. Ratti, and W. Weise, Phys. Rev. D\textbf{75}, 034007 (2007)
%\bibitem{Ghosh:2007wy} S. K. Ghosh, T. K. Mukherjee, M. G. Mustafa, and R. Ray, Phys. Rev. D77, 094024 (2008).
%\bibitem{Fu:2007xc} W.-j. Fu, Z. Zhang, and Y.-x. Liu, Phys. Rev. D\textbf{77}, 014006 (2008).
%\bibitem{Costa:2008dp} P. Costa, M. C. Ruivo, C. A. de Sousa, H. Hansen, and W. M. Alberico, Phys. Rev. D\textbf{79}, 116003 (2009).
\bibitem{Fukushima:2008wg}  K. Fukushima, Phys. Rev. D\textbf{77}, 114028 (2008). 
%\bibitem{Bhattacharyya:2010wp} A. Bhattacharyya, P. Deb, S. K. Ghosh, and R. Ray, Phys. Rev. D\textbf{82}, 014021 (2010).
\bibitem{Moreira:2010bx} J. Moreira, B. Hiller, A. Osipov, and A. Blin, Int.J.Mod.Phys. A\textbf{27}, 1250060 (2012).
%\bibitem{Stiele:2016cfs}  R. Stiele and J. Schaffner-Bielich, Phys. Rev. D\textbf{93},094014 (2016).
%\bibitem{Bhattacharyya:2016jsn} A. Bhattacharyya, S. K. Ghosh, S. Maity, S. Raha, R. Ray, K. Saha, and S. Upadhaya, Phys. Rev. D95, 054005 (2017).
\bibitem{Ghosh:2007wy} S. K. Ghosh, T. K. Mukherjee, M. G. Mustafa, and R. Ray, Phys. Rev. D\textbf{77}, 094024 (2008).
\bibitem{Bhattacharyya:2016jsn}  A. Bhattacharyya, S. K. Ghosh, S. Maity, S. Raha, R. Ray, K. Saha, and S. Upadhaya, Phys. Rev. D\textbf{95}, 054005 (2017).
\bibitem{Borsanyi:2011sw} S. Borsanyi, Z. Fodor, S. D. Katz, S. Krieg, C. Ratti, and K. Szabo, JHEP 01, 138 (2012)
\bibitem{Bazavov:2012jq} A. Bazavov et al. (HotQCD), Phys. Rev. D86, 034509 (2012).
\bibitem{Borsanyi:2013bia} S. Borsanyi, Z. Fodor, C. Hoelbling, S. D. Katz, S. Krieg, and K. K. Szabo, Phys. Lett. B\textbf{730}, 99 (2014).
\bibitem{Bazavov:2014pvz} A. Bazavov et al. (HotQCD), Phys. Rev. D\textbf{90}, 094503 (2014).
\bibitem{Moreira:2018xsp}J. Moreira, J. Morais, B. Hiller, A. A. Osipov, A. H. Blin, 	Phys. Rev. D\textbf{98}, 074010 (2018)
\end{thebibliography}

\end{document}